\documentclass[aps,prb,showpacs,showkeys,preprintnumbers,amsmath,amssymb]{revtex4-2}

\usepackage[utf8]{inputenc}
\usepackage{graphicx}
\graphicspath{{Figures/}}
\usepackage{hyperref}
\usepackage{units}
\usepackage{amsmath,amssymb}
\usepackage{textgreek}
\usepackage{multirow}
\usepackage{rotating}
\usepackage{color}
\usepackage{ulem}

\begin{document}
\title{Graphene-capped bismuthene on SiC as a platform for correlated quantum spin Hall edge states}
\author{Huu Thoai Ngo$^{1}$} 
\author{Zamin Mamiyev$^{1}$} 
\author{Niclas Tilgner$^{1,2}$} 
\author{Andres David Peña Unigarro$^{1,2}$} 
\author{Sibylle Gemming$^{1,2}$} 
\author{Thomas Seyller$^{1,2}$} 
\author{Christoph Tegenkamp$^{1}$} \email{christoph.tegenkamp@physik.tu-chemnitz.de}
\address{$^{1}$Institut f\"ur Physik, Technische Universit\"at Chemnitz, Reichenhainer Str. 70, D-09126 Chemnitz, Germany.}
\address{$^{2}$Center for Materials, Architectures and Integration of Nanomembranes (MAIN), Chemnitz, Germany.}

\date{\today}
\begin{abstract}
Epitaxial bismuthene on SiC(0001) hosts symmetry-protected metallic edge states within a large bulk band gap, establishing it as a promising two-dimensional topological insulator for high-temperature quantum spin Hall (QSH) transport. Here we realize bismuthene islands by intercalating Bi beneath zero-layer graphene on SiC(0001) followed by hydrogen treatment, yielding well-defined edges with controlled terminations. Spectroscopic measurements demonstrate that the edge states reside inside the bulk band gap and remain charge neutral. The graphene overlayer interacts only weakly with the bismuthene, preserving its topological character while providing environmental protection. Notably, the one-dimensional edge channels exhibit signatures of enhanced electronic correlations relative to freestanding bismuthene, suggesting proximity-induced modification of the QSH edge physics. These results establish graphene-capped bismuthene as a robust and tunable platform for correlated quantum spin Hall states.
\end{abstract}

{}

\maketitle

\section{Introduction}
Two-dimensional (2D) materials provide a versatile platform for emergent quantum phenomena, including dimen\-sionality-dependent electronic structure \cite{Kim@2024, Quan@2024}, symmetry breaking \cite{He@2021, Ngo@2026}, and non-trivial topology \cite{Qi@2011}, absent in bulk systems. Among them, 2D topological insulators (TIs) host symmetry-protected helical metallic edge states within a bulk band gap, giving rise to the quantum spin Hall (QSH) effect \cite{L.Kane@2005, Bernevig@2006}. These edge channels enable dissipationless transport \cite{Meng@2024} and are robust against non-magnetic disorder, offering prospects novel concepts in spintronics  and topological quantum computing \cite{Mellnik@2014,Wang@2015,Aasen@2016}. 

The quantum spin Hall (QSH) effect was first proposed by Kane and Mele in graphene with spin–orbit coupling \cite{L.Kane@2005}. Since then, it has been realized in diverse two-dimensional topological insulators, including group-IV monolayers (silicene \cite{Liu@2011}, germanene \cite{Bampoulis@2023}, stanene \cite{Deng@2018}), transition-metal dichalcogenides (1T'-$\rm WTe_2$ \cite{Tang@2017}, 1T'-$\rm MoS_2$ \cite{Katsuragawa@2020}), and even layered compounds such as $\rm Bi_4Br_2I_2$ \cite{Zhong@2023} and $\rm ZrTe_5$ \cite{Xu@2024}. 

Compared to these, bismuthene is a monoelemental QSH system with a large band gap that preserves its edge-state conductance even at elevated temperatures \cite{Lodge@2021}. It can be epitaxially grown with atomic precision on H-passivated SiC(0001) substrates \cite{F.Reis@2017, Stühler@2022}, offering a scalable and CMOS-compatible route toward device integration. Moreover, ambient protection of bismuthene has been achieved when it forms via Bi intercalation at the zero-layer graphene (ZLG)/SiC interface \cite{Gehrig@2025, Tilgner@2025}. However, the extent to which the edge-state properties are preserved or modified under such encapsulation remains unclear.

Beyond the QSH effect, the one-dimensional (1D) helical edge states in bismuthene host strong electron–electron interactions that invalidate the conventional Fermi-liquid picture \cite{R.Mélin@1994}. Instead, their low-energy excitations are described by a Tomonaga–Luttinger liquid (TLL) with spin–momentum locking and a power-law suppression of the local density of states (LDOS). Epitaxial bismuthene on H–SiC has been shown to exhibit TLL behavior at both armchair and zigzag edges, manifested by a zero-bias anomaly (ZBA) in tunneling spectra \cite{Stuhler@2020, Stühler@2022}. Whether this TLL behavior persists in bismuthene intercalated beneath zero-layer graphene (ZLG) remains unknown, as the modified dielectric environment, screening, and interface coupling introduced by intercalation may strongly alter electron–electron interactions. Resolving this question is crucial for understanding the robustness of topological edge states in intercalated bismuthene.
 
In this work, we investigated bismuthene intercalated beneath zero-layer graphene on SiC(0001) using low-temperature scanning tunneling microscopy/spectroscopy (LT-STM/STS), spot-profile analysis low-energy electron diffraction (SPA-LEED), and density functional theory (DFT) calculations. STM imaging reveals that Bi intercalation leads to formation of nanoscale bismuthene islands with well-defined armchair edges. Spatially resolved STS measurements show a bulk band gap and mid-gap metallic edge states. Remarkably, tunneling spectra at the edges display a pronounced suppression of the local density of states (LDOS) near the Fermi level, evidencing Tomonaga–Luttinger liquid (TLL) behavior, where electronic correlations in the one-dimensional edge channels are enhanced by the weakly interacting, proximitized graphene overlayer.


\section{Results}
\subsection{Structure and morphology of 2D-Bi islands}
\begin{figure*}[ht]
	\centering
	\includegraphics[width=0.8\linewidth]{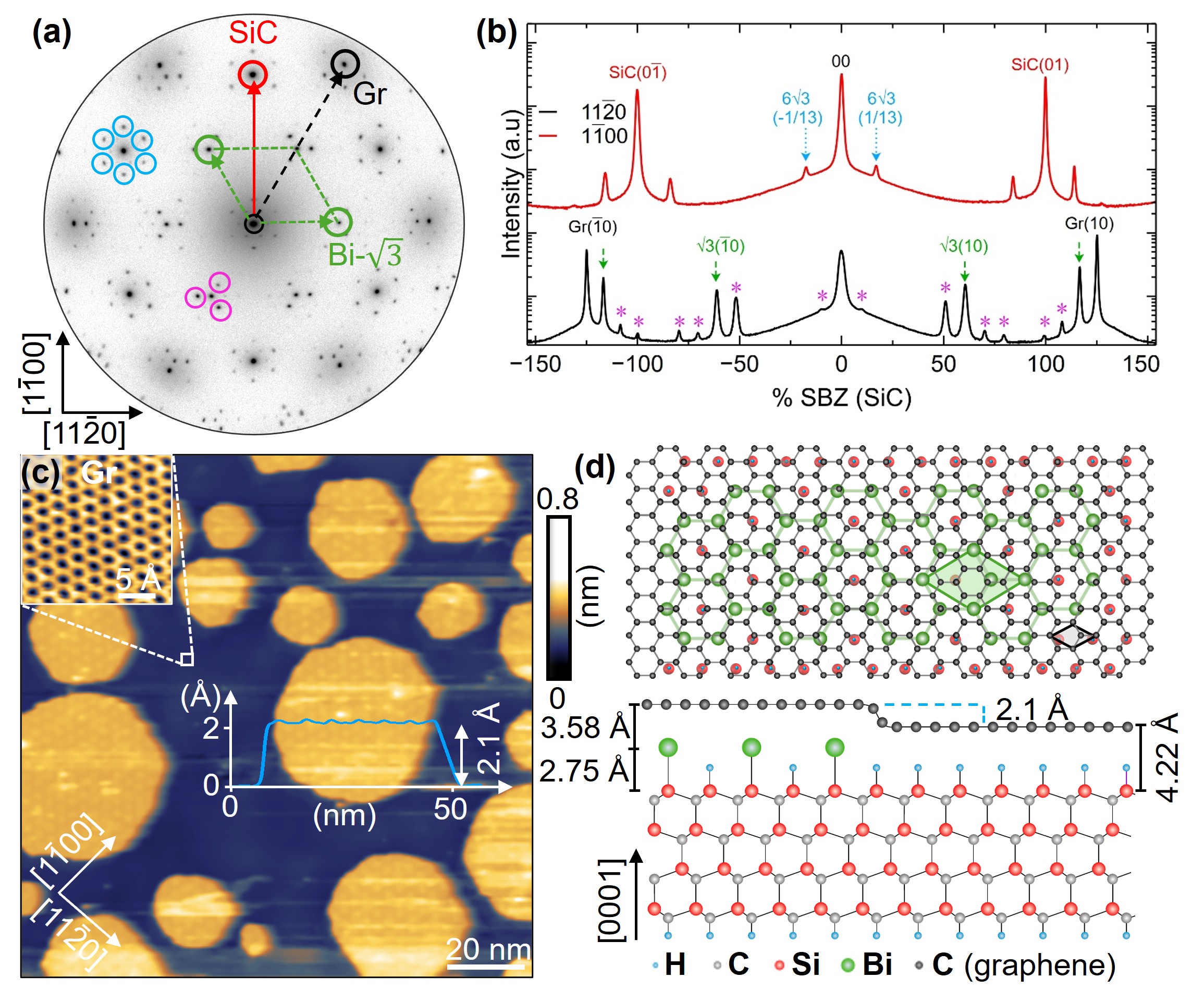}
	\caption{\textbf{Intercalation and hydrogenation of Bi below ZLG.} (a) SPA-LEED pattern taken at E=167~eV of Bi-intercalated ZLG/SiC showing the graphene (black circle), Bi-$(\sqrt{3}\times\sqrt{3})R30^{\circ}$ (green circle), SiC (red circle), $(6\sqrt{3} \times 6\sqrt{3})R30^{\circ}$ (purple circle) and ($6\times6$ (blue circle) periodicity. (b) Line profiles along two crystallographic directions. All diffraction experiments were performed at T = 300~K. (c) Large-scale STM topographic image (+1~V, 450~pA, 4.5~K)  showing intercalated bismuthene islands (EG/Bi/SiC) surrounded by H-intercalated ZLG area. Inset: STM image showing atomically resolved HQFMLG  recorded at +1.8 V, and 300 pA, T = 4.5 K. The line scan across the bismuthene islands reveals an apparent height of 2.1\AA. (d) Ball and stick model of bismuthene- and H-intercalated ZLG on SiC: top view (top) and side view (bottom). 
\label{Figure1}} 
\end{figure*}

The high-resolution diffraction pattern shown in Fig.~\ref{Figure1}a reveals the characteristic diffraction spots of SiC(0001), graphene, and bismuthene, as well as the ($6\sqrt{3}\times 6\sqrt{3}$) Moiré spots of graphene relative to the SiC substrate.
The line scans in panel b show not only the first- and higher-order spots of the bismuthene $(\sqrt{3}\times\sqrt{3})R30^{\circ}$ reconstruction, reminiscent of the formation of well-ordered bismuthene phase, but also the bell-shaped background indicative of a fully delaminated ZLG and its transformation into a quasi-free-standing monolayer graphene  \cite{Chen@2019, Mamiyev2022, Tilgner2025}.
 
The large-scale STM image (Fig.~\ref{Figure1}c) fully supports these findings and additionally reveals the formation of numerous islands with sizes on the order of 10–50 nm (SI, Fig.~\ref{Figure1}g ). The islands reveal a Moir\'e structure, arising from the lattice mismatch between the SiC ($a_{\mathrm{SiC}}$ = 3.08 \AA), bismuthene, ($a_{\mathrm{Bi}}$ = $5.32$ \AA) and graphene ($a_{\mathrm{Gr}}$ = 2.46 \AA). Prior to the formation of the bismuthene phase, a  $\beta$ Bi phase with -$(\sqrt{3} \times \sqrt{3})R30^{\circ}$ symmetry is formed. Thereby, Bi atoms occupy  $T_4$ hollow sites of a SiC surface and form three Bi-Si bonds, saturating all Si dangling bonds and preventing planar Bi-Bi hybridization \cite{Sohn2021, Wolff_2024, Tilgner@2025}. To form finally bismuthene, the $\beta$- Bi phase was hydrogenated. Hydrogen atoms preferentially passivate Si dangling bonds; thus, the  Bi atoms are shifted to the $T_1$ positions above the Si atoms, leading finally to the formation of bismuthene  \cite{Tilgner@2025}. The graphene atop is  a quasi-free MLG (BiQFMLG in the following), separated at a distance of 3.58 \AA\ from the intercalated bismuthene layer. In the bismuthene phase, each Bi atom is covalently bonded only to a Si atom of the substrate, thus enabling planar Bi-Bi hybridization and stabilizing the bismuthene honeycomb lattice, as depicted in Fig.~\ref{Figure1}d). Moreover,  the hydrogenation process leads to the formation of hydrogen-intercalated areas, revealing a second type of quasi-free-standing monolayer graphene (HQFMLG) \cite{Riedl2009}, thus stabilizing certain edges of bismuthene islands and ensuring a full carpeting with a graphene layer. 
This is obvious from the line scan shown in Fig.~\ref{Figure1}c). The apparent height difference of $\sim 2.1\,\text{\AA}$ fits nicely to the heights reported for bismuthene and H- intercalated ZLG of $\sim 6.33\,\text{\AA}$  and $\sim 4.22\,\text{\AA}$, respectively (SI, Fig.~\ref{Figure1}h) \cite{Tilgner_2025_b,Sforzini@2015}. 

The observation that predominantly islands are formed may provide insights into the coverage of the individual phases. Starting from the $\alpha$-Bi phase, which corresponds to a monolayer coverage, desorption leads to the formation of a $\beta$-Bi precursor phase. Upon hydrogenation, this phase transforms into bismuthene. The coverage of the bismuthene honeycomb phase corresponds to two-thirds of a monolayer. Assuming that the $\beta$-Bi phase indeed exhibits a coverage of one-third of a monolayer, as also suggested in the literature \cite{Sohn2021}, island formation is indeed to be expected.
Compared to previous studies, in which the edges of the 2D QSH system were often located close to SiC substrate steps \cite{F.Reis@2017, Stuhler@2020, Syperek@2022}, we realized encapsulated bismuthene islands on flat SiC terraces, which facilitates the analysis of the edges and their associated edge states as we will show in the following.\\

\begin{figure*}[tb]
	\centering
	\includegraphics[width=0.8\linewidth]{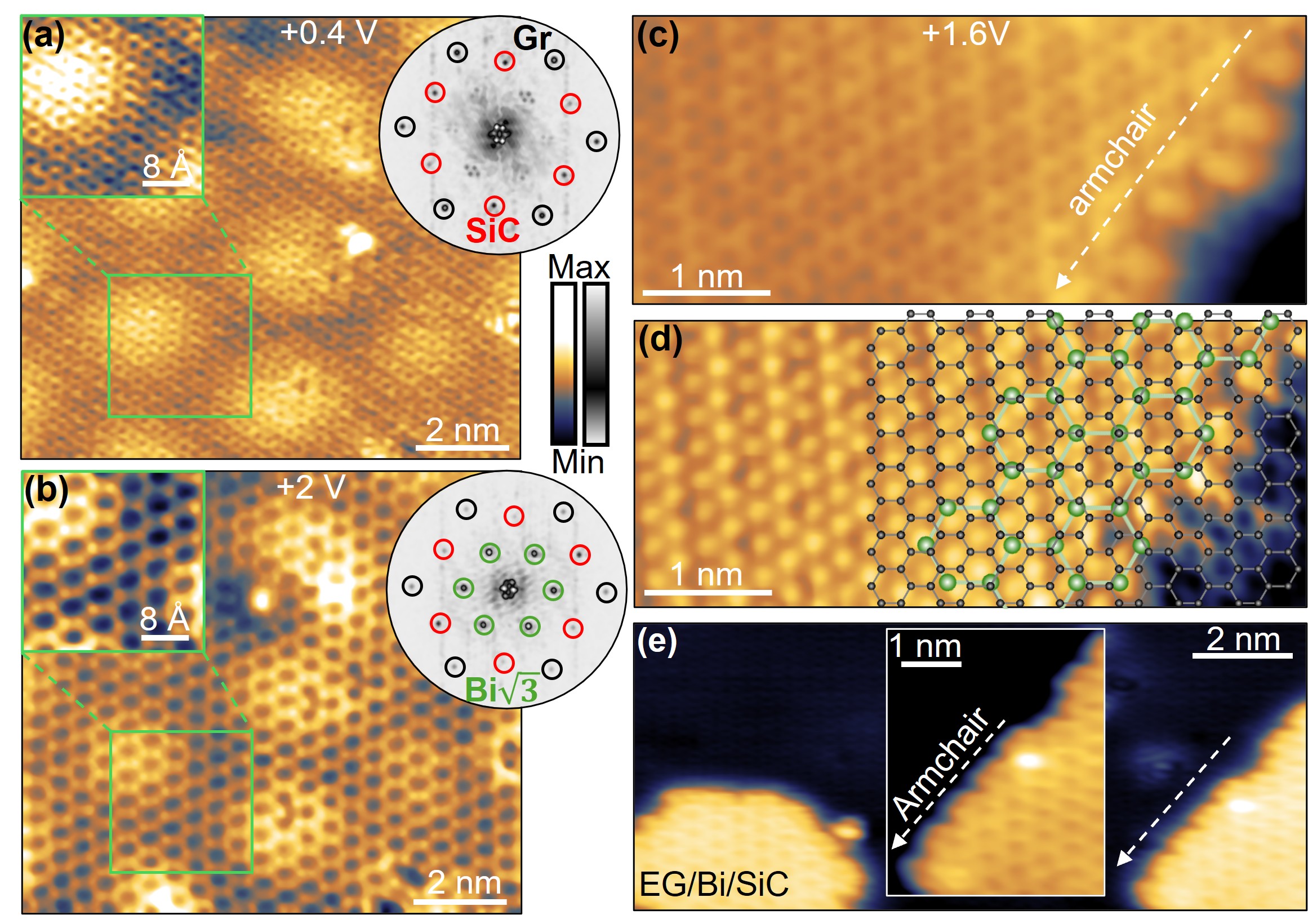}
	\caption{\textbf{Bismuthene structure and edges.} (a-b) STM topographic and FFT images of a bismuthene-intercalated ZLG island  measured at two different bias voltages: $V_b$ = +0.4 V (a), and $V_b$ = +2 V (b). Inset: Zoomed-in STM images (left) of the regions outlined by the green rectangles, showing the honeycomb lattice of graphene and bismuthene. The corresponding FFT images are presented in Fig.~\ref{Figure1}a)-b) (right).  (c-d) STM topographic (c) image and current map (d) of an intercalated bismuthene island exhibiting the armchair termination of bismuthene (+2 V, 450 pA). Schematic illustration of the graphene (black) and Bi-honeycomb lattice (green) is superimposed on the STM image. (e) STM topography of a small region exhibiting two intercalated bismuthene islands (EG/Bi/SiC) (+3 V, 250 pA). Inset: Zoomed-in STM image of the armchair edge. The STM images were acquired at 4.5 K.
\label{Figure2}} 
\end{figure*}

To further corroborate these findings and resolve the sublattices of both 2D heterosystems, the sample was examined by bias-dependent STM measurements. Bismuthene is a topological insulator with a finite bulk band gap, in contrast to the semi-metallic character of the graphene overlayer. This distinction allows the lattice of each material layer to be probed selectively by adjusting the bias voltage in STM measurements. Figs.~\ref{Figure2}a)-b) show constant-current STM images of the same region, where each of the two lattices can be probed by selecting bias voltages. Indeed, at $V_b$ = +0.4 V, the STM image shows only the graphene honeycomb lattice. The Bi honeycomb lattice (bismuthene) appears at a bias voltage of +2 V, mostly probing the states of the bismuthene bulk bands, as also confirmed by STS (see below). 
The corresponding FFT images are presented in Fig.~\ref{Figure2}a)-b) as well, revealing the Fourier coefficients of the graphene, bismuthene, and SiC(1$\times$1) lattices marked by the black, green, and red circles, respectively. The FFT image in Fig.~\ref{Figure2}a) ($V_b$ = +0.4 V) exhibits the intense spots of graphene in addition to the SiC ($1\times1$) and Moir\'e pattern, indicating the formation of QFMLG upon intercalation. Meanwhile, the bismuthene layer is directly probed at $V_b$ = +2 V, giving rise to intense spots of the $(\sqrt{3} \times \sqrt{3})R30^{\circ}$ lattice in Fig.~\ref{Figure2}b), in good agreement with the SPA-LEED pattern (SI, Fig.~\ref{Figure1}b-d). 

In the following, we focus on elucidating the edge structure of the intercalated bismuthene islands. This interpretation is supported by the atomic-resolution STM image (Figs.~\ref{Figure2}c-e), acquired by scanning over a small region. The STM image in Fig.~\ref{Figure2}c) reveals the interface structure of the island, where both lattices are resolved at the atomic scale. Remarkably, the island boundary corresponds to an armchair termination of the intercalated bismuthene layer \cite{Sun@2022}. The graphene overlayer remains continuous across the armchair-terminated boundary, exposing alternating Bi dimers at the edge, as evidenced by the current mapping image in Fig.~\ref{Figure2}d). This observation agrees well with the STM topography in Fig.~\ref{Figure2}e), where the alternating Bi dimers are clearly observed at the armchair edge.

For bismuthene islands grown on SiC(0001), predominantly armchair-type edges are observed, even in the intercalated case \cite{F.Reis@2017, Stuhler@2020}. This preference has been attributed to step-induced edge formation on the SiC substrate \cite{Stühler@2022}. In our case, armchair edges are found in nearly all islands ($\approx$75\%) regardless of the presence of SiC step edges (SI, Fig.~\ref{Figure2}e), suggesting an intrinsic energetic preference. The measured periodicity along the edges, $\sqrt{3}a_{Bi}=3a_{SiC}=9.24$~\AA, agrees with the expected lattice match, excluding relaxation effects. Armchair edges also appear chemically more stable than zigzag ones, which are prone to degradation during hydrogen post-treatment or due to nearby adsorbed hydrogen (SI, Fig.~S2b). Notably, such edge terminations have recently been shown to influence the topological character of bismuthene \cite{Wang2025}.

\subsection{Electronic bulk properties of bismuthene under cover}
\begin{figure*}[tb]
	\centering
	\includegraphics[width=0.7
    \linewidth]{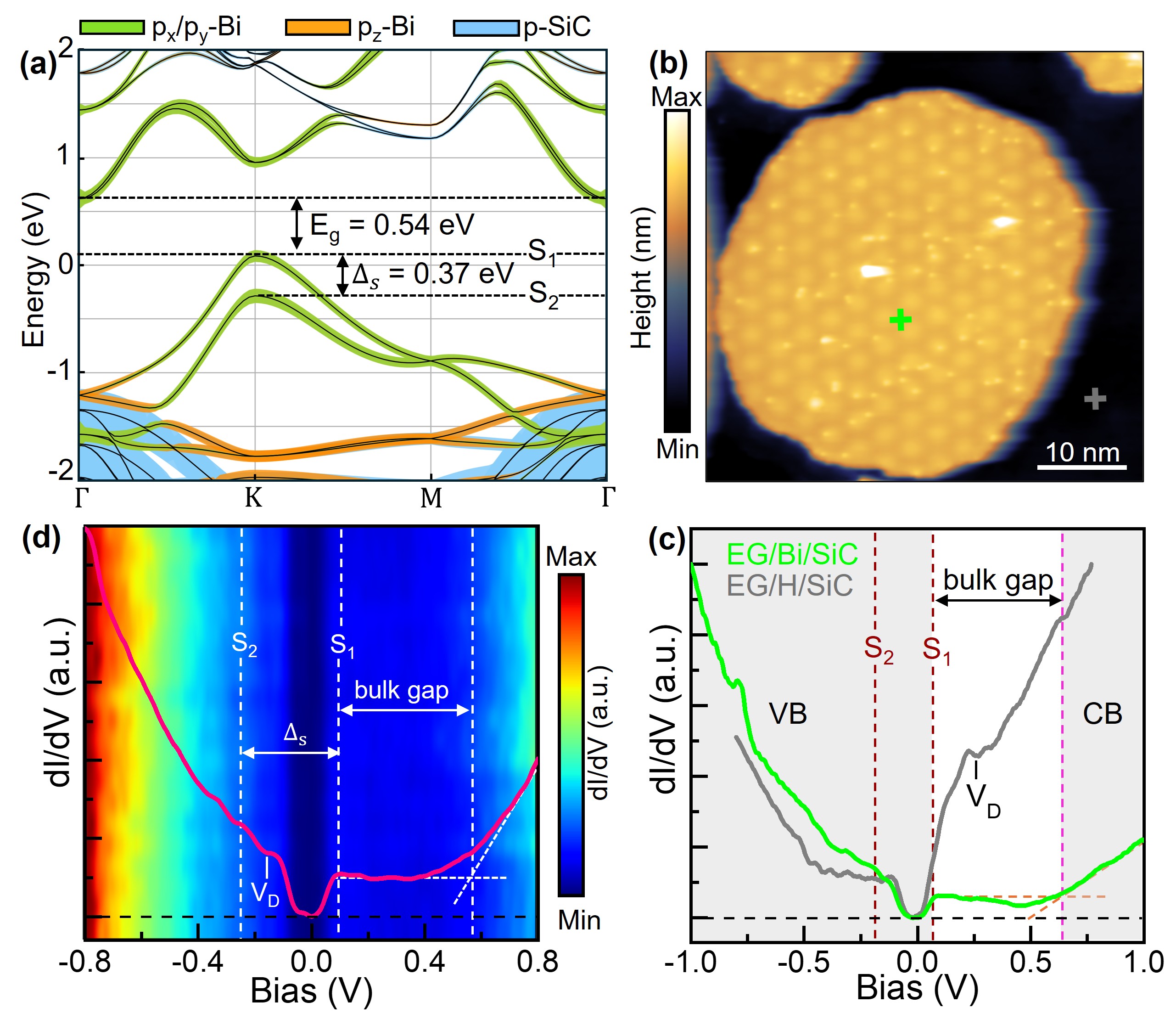}
	\caption{\textbf{Bulk electronic structure of bismuthene.} (a) DFT calculations for the band structure of bismuthene/H-passivated SiC. The indirect band gap of bismuthene is denoted as $E_g$, while $\Delta_s$ represents the energy spacing between the $S_1$ and $S_2$  bands of the valence band (VB). (b) STM topography of intercalated bismuthene islands (EG/Bi/SiC) surrounded by H-intercalated ZLG region (EG/H/SiC) (+0.8 V, 450 pA). (c) Tunneling dI/dV spectra (STS) recorded at two different positions marked by the green and gray crosses as indicated in Fig.~\ref{Figure3}a). Tunneling conditions are $V_b$ = +1 V, $I_t$ = 450 pA (green), and $V_b$ = +0.8 V, $I_t$ = 400 pA (gray). Two tangential lines  (orange) were drawn to the green STS curve to determine the conduction band (CB) edge. The Dirac point is denoted as $V_D$. (d) Average dI/dV spectrum measured at the island interior (green cross) on the top of 20 consecutive dI/dV curves concatenated into a color map ($V_b$ = +0.8 V, $I_t$ = 450 pA). In the STS map, $S_1$ and $S_2$ represent the top and next energy levels of the VB, respectively. The black-dashed line represents the zero dI/dV intensity. The STM/STS results were acquired at 4.5 K. 
\label{Figure3}} 
\end{figure*}

Recent ARPES results showed  that the graphene overlayer interacts only weakly with the underlying bismuthene layer, acting primarily as a chemically inert and protective cap \cite{Tilgner@2025, Gehrig@2025}. This weak interaction suggests that the graphene layer does not significantly modify the intrinsic electronic structure of the intercalated bismuthene layer. Motivated by this picture, we first examine the bulk electronic structure of bismuthene on H-passivated SiC substrate by means of density functional theory (DFT) (SI, Fig. S6a). As shown in Fig.~\ref{Figure3}a, the valence band (VB) near the Fermi level, primarily composed of Bi $p_x/p_y$ orbitals, splits into two bands ($S_1$ and $S_2$) with an energy separation of 0.37 eV induced by the strong spin–orbit coupling of Bi. Notably,  the $S_1$ band lies slightly above the Fermi level as a result of substrate-induced p-doping, consistent with ARPES measurements \cite{Tilgner@2025}. Furthermore, the conduction-band minimum at the $\Gamma$ point gives rise to an indirect band gap of 0.54~eV.  DFT calculations for the heterostructure including the graphene layer were also performed using an approximate $(2\times2)$ graphene supercell on bismuthene/($\sqrt{3}\times \sqrt{3}$)-SiC (SI, Fig. SI6b-e). The results reveal n-type doping of graphene and negligible interlayer hybridization, confirming weak van der Waals interaction between the layers.

\begin{figure*}[t]
	\centering
	\includegraphics[width=0.8\linewidth]{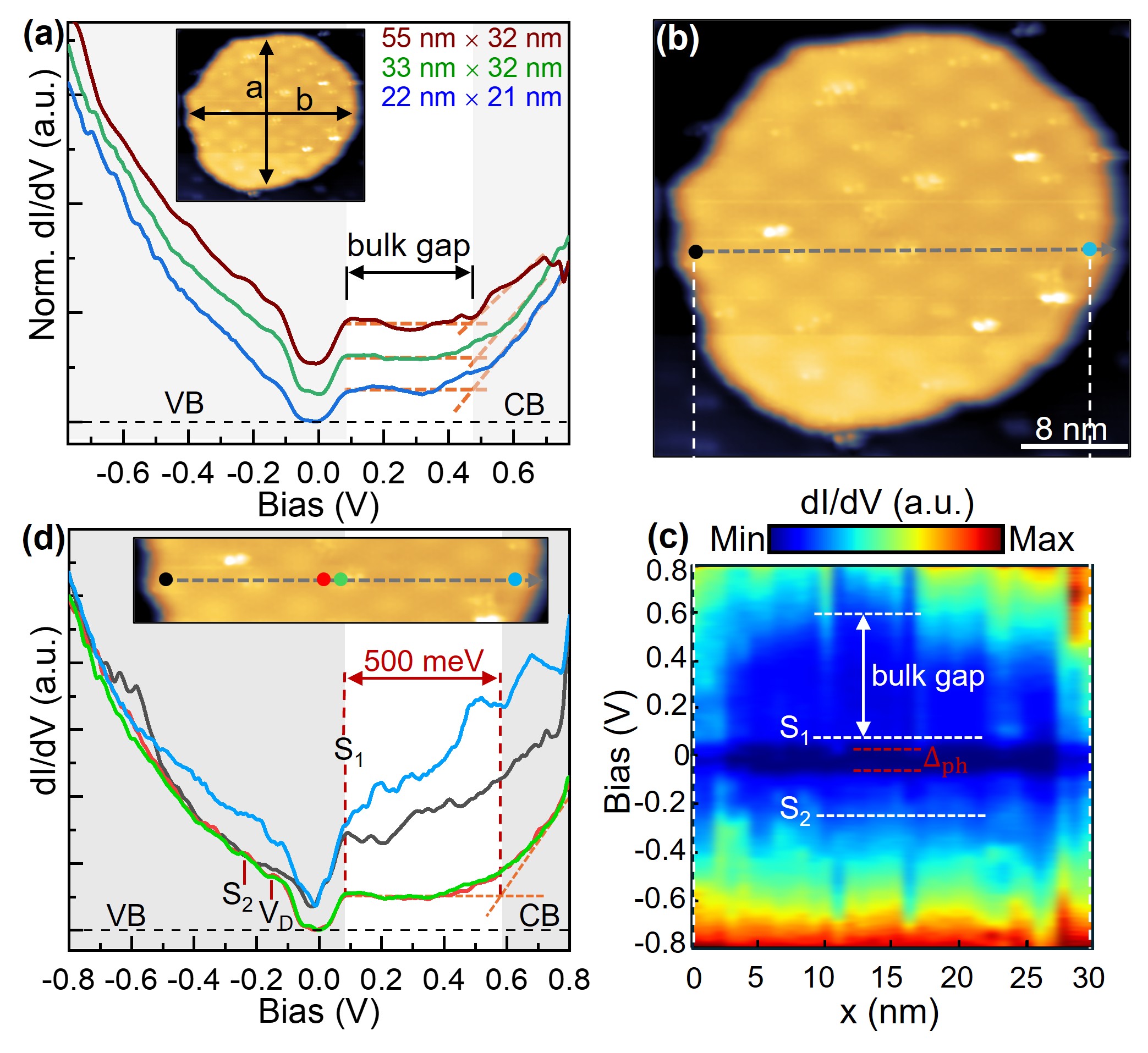}
	\caption{\textbf{Size- and side-effects of bismuthene islands.} (a) Tunneling dI/dV spectra of three intercalated bismuthene islands with different lateral sizes. The spectra were acquired at an interior position of each island. Island 1 ($22 \times 21\,\mathrm{nm}^2$): $V_b$ = 0.8 V, $I_t$ = 500 pA. Island 2 ($33 \times 32\,\mathrm{nm}^2$): $V_b$ = 0.8~V, $I_t$ = 450~pA. Island 3 ($55 \times 32\,\mathrm{nm}^2$): $V_b$ = 0.8 V, $I_t$ = 370 pA. The spectra are shifted for better visibility. (b) STM topography of an intercalated bismuthene island measured at +1.5 V and 400 pA. (c) Line scan of dI/dV spectra ($V_b$ = +0.8 V, $I_t$ = 450 pA) measured along the gray dashed line shown in Fig.~\ref{Figure4}b. The black and blue circles represent the starting and ending positions. The Dirac point and the phonon-induced gap are denoted as $V_D$, and $\Delta_{\mathrm{ph}}$. (d) Representative dI/dV spectra were extracted from the positions marked by the colored circles, shown in the inserted image (cut off from Fig.~\ref{Figure4}b). Two tangential lines (orange) were drawn to the green and red STS curves to determine the CB edge. The black dashed lines represent the zero dI/dV intensity. The STM/STS data were measured at 4.5 K.
\label{Figure4}} 
\end{figure*}
 
To further elucidate the electronic structure, scanning tunneling spectroscopy (STS) was employed to probe the local density of states (LDOS) of the intercalated bismuthene islands. Figure \ref{Figure3}c shows representative dI/dV spectra acquired at the island interior (Fig. \ref{Figure3}b, green cross) and on the H-intercalated ZLG region (Fig. \ref{Figure3}b, gray cross). For EG/Bi/SiC, the spectrum measured on the island (green) displays distinct features, including the upper valence band state ($S_1$) at +0.1 V and a nearly constant DOS region above $S_1$. The onset near +0.5 eV corresponds to the bulk conduction band of bismuthene and agrees well with DFT calculations. The DOS within this gap, however, is not completely suppressed, reflecting the contribution of the overlying graphene layer. The STS curve reveals a suppression of the DOS at $E_F$, accompanied by an apparent gap. This characteristic feature arises from inelastic tunneling processes mediated by a phonon that supplies the necessary momentum for transitions into the graphene $\pi$-states at the K-point \cite{Zhang@2008}. The observation of this gap is a well-established hallmark of freestanding, structurally intact graphene.

We also measured the $dI/dV$ curve obtained on HQFMLG (Fig.~\ref{Figure3}c). Apart from the inelastic tunneling gap, which is also expected for HQFMLG, the spectrum exhibits distinct features. In particular, the small dip around $+0.25~\text{eV}$ can be associated with the Dirac point $E_D$ \cite{Goler@2013}. The observed $p$-type doping for the only H-intercalated ZLG areas is consistent with ARPES results \cite{Riedl2009, Tilgner@2025}.

The results presented in Fig.~\ref{Figure3}(d) show a total of 20 spectra taken at an interior position of the intercalated bismuthene island, demonstrating the reproducibility and robustness of the above discussed spectral fingerprints. The pink curve represents the averaged spectrum, enabling a clearer identification of the $S_1$ and $S_2$ states, as well as the SOC splitting ($\Delta_s$ = 0.34 eV), in good agreement with the DFT calculations.  The dip  at -0.16 V attributed to the Dirac point ($V_D$) for the n-doped graphene on top of the bismuthene layer \cite{Tilgner@2025}. 

\begin{figure*}[tb]
	\centering
	\includegraphics[width=0.98\linewidth]{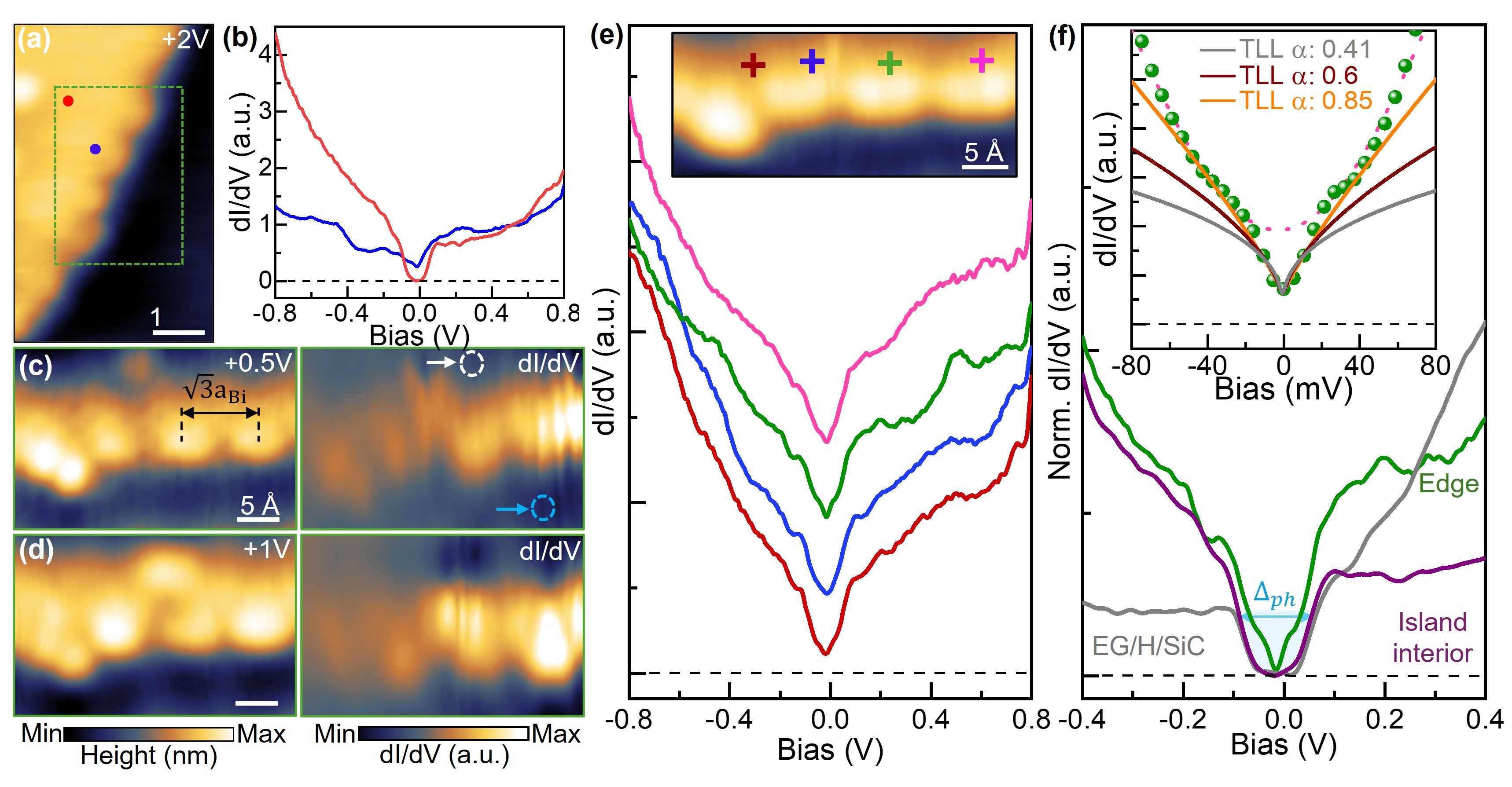}
	\caption{ \textbf{Spectroscopy of the edge states.} (a) STM topography of an intercalated bismuthene island boundary (+2 V and 450 pA).  (b) dI/dV spectra (STS) recorded at two different positions marked by the red and blue circles in Fig.~\ref{Figure5}a. (c-d) STM topographic (left) and conductance (dI/dV) mapping images (right) of a small region outlined by the green dashed rectangle in Fig.~\ref{Figure5}) measured at (c) +0.5 V, and (d) +1 V. The white and blue arrows indicate the island interior and EG/H/SiC, respectively. (e) STS recorded at various positions (crosses) along the armchair edge. Inset: STM topography (+2 V, 450 pA) shows the probed positions. (f) STS measured at a H-intercalated ZLG region compared to the STS recorded at an island interior and boundary (armchair edge). For better comparison, the spectra were shifted to zero conductance. The phonon-induced gap ($\Delta_{\mathrm{ph}}$) is outlined by a blue rectangle. Inset: Zoomed-in STS of the armchair edge (green) was fitted using the TLL model with $\alpha$ = 0.41 (gray), $\alpha$ = 0.6 (brown), and $\alpha$ = 0.85 (orange). The STS data (green data points) was shifted on the energy scale to zero. The pink dashed curve indicates the phonon-induced gap contributed by the graphene overlayer. The black dashed line represents the zero dI/dV intensity. All dI/dV spectra measured at $V_b$ = +0.8 V, $I_t$ = 450 pA, T = 4.5 K. 
\label{Figure5}} 
\end{figure*}
 
\subsection{Island size and edge effects of bismuthene islands}
We have shown above a weak electronic coupling between graphene and bismuthene, allowing us to study the intrinsic electronic structure of the intercalated bismuthene layer. The formation of islands with different sizes naturally raises the question of how this affects the bulk electronic bands of the intercalated bismuthene. The $dI/dV$ spectra shown in Fig.~\ref{Figure4}(a) for three different islands display very similar spectral features. Although the local DOS varies slightly, no significant modifications are observed in the band structure measured at the interior of each island. This indicates that the bulk band gap of bismuthene remains essentially insensitive to the island’s lateral dimensions, as expected for 2D topological insulators \cite{Hohenadler@2013}.

To analyze the edges and demonstrate the existence of edge states, in order to be termed a QSH system, we measured $dI/dV$ spectra across a capped bismuthene island, shown in Fig.~\ref{Figure4}b). The corresponding spectra are displayed as an intensity map in panel (c). In the center region, we mainly observe the same spectral features as discussed  in context of Fig.~\ref{Figure3}d).

Remarkably, the bismuthene band gap gradually decreases and eventually vanishes as the STM tip approaches the island boundaries. The phonon-induced gap of graphene appears to persist but is superimposed by a finite intensity within the former gap region. Figure~\ref{Figure4}d) shows representative spectra taken at the bulk and edge positions. Instead of the characteristic suppression of intensity at positive bias voltages, a distinct V-shaped feature emerges. 
Most importantly, the overall conductance ($dI/dV$ intensity) measured at the island boundaries is higher than that of the island interior, confirming the metallic character of the bismuthene edges. 
These $dI/dV$ results unambiguously demonstrate the presence of metallic edge channels in the intercalated bismuthene islands, which are a prerequisite for the QSH effect, despite the presence of the graphene overlayer \cite{F.Reis@2017, Stuhler@2020}.

\subsection{Edge states along bismuthene islands}
We next examine dI/dV conductance maps of a distinct armchair edge, whose morphology and site-resolved electronic structure are shown in Figs. \ref{Figure5}a and b. Figures \ref{Figure5}c–d present constant-current STM images (left) alongside the corresponding dI/dV maps (right) acquired at various bias voltages. The STM images reveal an enhanced tunneling signal with a characteristic periodicity of $\sqrt{3}a_{\text{Bi}} = 0.92$ nm, consistent with armchair edges in pristine bismuthene \cite{Stuhler@2020}. In the dI/dV maps, a pronounced and spatially localized enhancement of tunneling conductance is observed along the edge, in stark contrast to the suppressed conductance within the island interior (white arrow) and the HQFMLG region (blue arrow). This clear contrast indicates the metallic nature of the edge states.

To further characterize the helical edge states,  tunneling spectra were measured at different positions along the investigated armchair edge (cf. Fig.~\ref{Figure5}e and inset). The spectra exhibit the characteristic V-shaped line shape with a finite offset at $E_F$, in agreement with previous measurements on uncovered bismuthene edge states \cite{F.Reis@2017}.
Most interestingly, all spectra show, in addition, a small dip at low energies. Compared to freestanding, i.e., non-intercalated bismuthene, such a zero-bias suppression was associated with a TLL-behavior due to electronic correlations within edge states (SI, Fig.~\ref{Figure5}) \cite{F.Reis@2017, Stuhler@2020, Li@2016}.
For our intercalated bismuthene islands, the intrinsic TLL signature is only faintly visible at small bias voltages and is partially suppressed by the inelastic gap of the graphene overlayer, as becomes evident from the direct comparison with the $dI/dV$ signal acquired at bulk positions (Fig.~\ref{Figure5}f). 

This limitation precludes a detailed quantitative analysis of the spectra. Instead, we interpret the observed behavior within the framework of a TLL-model using different Luttinger parameters $\alpha$, as shown in the inset of Fig.~\ref{Figure5}f. A value of $\alpha = 0.41$, previously reported for extended, uncovered armchair bismuthene edges \cite{Stuhler@2020}, reproduces the experimental density of states only within a narrow bias range (±10 mV). To capture the spectral line shape over a wider voltage interval, a larger exponent of $\alpha = 0.85$ is required (Fig.~\ref{Figure5}f, inset). This result indicates enhanced electron–electron interactions in the intercalated bismuthene heterostructure, likely arising from a modified dielectric screening by the SiC substrate and the graphene overlayer \cite{Jia2022}. Comparable large $\alpha$ values have been reported in other strongly correlated systems, such as 1T'-
$\rm WTe_2$/BLG \cite{Jia2022} and folded graphene edges \cite{Cai@2025}. These observations suggest that while the graphene overlayer does not alter the topologically non-trivial edge states of bismuthene, it promotes stronger electronic correlations within the system.

\section{Discussion}
In summary, we investigated the structural and electronic properties of bismuthene intercalated ZLG using LT-STM/STS, SPA-LEED, and DFT calculations. The intercalation leads to the formation of a well-defined bismuthene lattice beneath epitaxial graphene, featuring distinct edge terminations, particularly the armchair edges of the bismuthene film. Spatially resolved STS measurements reveal that the island interior is characterized by a large bulk band gap of approximately 0.5~eV, whereas metallic edge states are observed at the island boundaries. The Dirac point coincides with the chemical potential, i.e., the edge states are effectively charge neutral. Moreover, the observed band warping toward the edges suggests that the edge states lie within the bulk band gap. 
The calculated band structure of the island interior agrees well with our experimental results, and the edge states show indications of electronic correlations. 
The detailed analysis of the edge spectra has shown that the signatures of electronic correlations in the 1D edge channels are tendentially more pronounced than in graphene-uncovered bismuthene islands. In this context, graphene not only protects the bismuthene from environmental influences but, through its screening properties, also appears to promote stronger electronic correlations in the one-dimensional channels, thus providing excellent conditions for transport measurements and even potential applications at elevated temperatures.

\section{Methods}
\textbf{Sample preparation:}
ZLG was epitaxially grown on a SiC(0001) substrate using the polymer-assisted sublimation growth, as described in detail elsewhere \cite{Kruskopf@2016, guse2025growth}. Following initial degassing and characterization, Bi deposition was performed in a dedicated UHV chamber with a base pressure of less than $1 \cdot 10^{-6}$\,Pa by evaporation from a Knudsen cell kept at $550^\circ$C for 120\,min. After transferring the sample to the analysis chamber (base pressure $< 2 \cdot 10^{-8}$\,Pa), Bi intercalation was achieved by annealing at $450^\circ$C for 30\,min. The sample temperature was monitored with an infrared pyrometer (emissivity 0.9). Transformation to the $\beta$-phase was achieved by Bi dilution at elevated temperatures ($950^\circ$C for 10\,min), as described in previous reports \cite{Sohn2021, Wolff_2024}. Bismuthene formation was accomplished by annealing the sample at $550^\circ$C for 90\,min in a dedicated contactless infrared heating system under 850\,mbar of H$_2$ at a flow rate of 0.9\,slm. Prior to STM measurements, the sample quality was verified through photoelectron spectroscopy, in comparison with previous studies \cite{Tilgner@2025, Gehrig@2025}. \\

\textbf{LT-STM and SPA-LEED measurements:}
After intercalation, the samples were immediately transferred via a UHV suitcase to the SPA-LEED system to examine the overall sample quality ($p = 10^{-9}\mathrm{Pa}$). The high-resolution diffraction profiles allow for the determination of surface roughness and the presence of graphene. Subsequently, the samples were transferred in situ into the STM chamber, where their structural and electronic properties were investigated at low temperatures (4.5 K and 77 K) under UHV conditions. PtIr tips were calibrated on Au(111) substrates prior to STM/STS measurements. STM measurements were carried out in constant-current mode. Tunneling $dI/dV$ spectroscopy (STS) was performed in constant-height mode using a lock-in amplifier with a modulation voltage of $V_{\mathrm{rms}} = 12~\mathrm{mV}$ and modulation frequencies of $f_{\mathrm{rms}} = 860~\mathrm{Hz}$ and $750~\mathrm{Hz}$.\\

\textbf{Computational details:}
The structural relaxations of the initial configurations and the calculation of the electronic properties, including the band structure and the density of states (DOS), were performed within density functional theory (DFT) as implemented in the ABINIT package \cite{gonze2009abinit,Gonze2020}. The exchange–correlation potential was treated within the generalized gradient approximation using the Perdew–Burke–Ernzerhof (PBE) functional \cite{perdew1996generalized}. Fully relativistic norm-conserving pseudopotentials were employed throughout. Structural relaxations were carried out using an $8\times8\times1$ Monkhorst–Pack $k$-point mesh until all atomic forces were below 0.0025 eV/Å. Dispersion interactions between layers were taken into account using the D3 van der Waals correction scheme \cite{Grimme2011}. For the self-consistent calculations, a total energy convergence threshold of $2.7\times10^{-8}$ eV and a plane-wave kinetic energy cutoff of 1088 eV were used.

The intercalated graphene structure was modeled using a commensurate supercell approach, in which a $(2\times2)$ graphene supercell corresponds to a $(\sqrt{3}\times\sqrt{3})$ SiC supercell. This commensurability was achieved by applying an in-plane strain of approximately 8\% to the graphene layer \cite{PhysRevLett.99.076802, PhysRevMaterials.5.074004}. At the interface between graphene and the SiC substrate, a bismuthene monolayer with $(\sqrt{3}\times\sqrt{3})$-SiC periodicity was introduced \cite{Tilgner2025, F.Reis@2017}. In this configuration, two of the three Si atoms in the topmost SiC surface layer were saturated with Bi atoms, while the remaining Si atom was passivated with hydrogen. The dangling bonds at the bottom surface of the SiC substrate were also passivated with hydrogen atoms. A vacuum spacing of 15 Å was included along the out-of-plane direction to avoid spurious interactions between periodic images.\\

\section*{acknowledgments}
We gratefully acknowledge financial support from the DFG through the Research Unit FOR5242 (project 449119662) and  Te386/25-1 (project 509747664).
%

\end{document}